\documentstyle[11pt,paspconf]{article}
\def\simless{\mathbin{\lower 3pt\hbox
   {$\rlap{\raise 5pt\hbox{$\char'074$}}\mathchar"7218$}}}   % < or of order
\def\simgreat{\mathbin{\lower 3pt\hbox
   {$\rlap{\raise 5pt\hbox{$\char'076$}}\mathchar"7218$}}}   % > or of order
\def\etal{{\rm et al.}}
\input psfig

\def\solmas{{M$_\odot$}}
\def\solm{{M_\odot}}
\def\be{\begin{equation}}
\def\ee{\end{equation}}
\def\macc{\dot M_*}
\def\racc{R_{\rm acc}}
\def\rbh{R_{\rm BH}}
\def\rroche{R_{\rm roche}}
\def\rs{R_*}
\def\ms{M_*}
\def\etal{et.al.}
\def\vel{v}
\def\menc{M_{\rm enc}}

\begin{document}

\title{Competitive Accretion in Clusters and the IMF}
\author{Ian A. Bonnell}
\affil{University of St Andrews, Physics and Astronomy, North Haugh, St Andrews, KY16 9SS, UK}

\begin{abstract}
Observations have revealed that most stars are born in
clusters.  As these clusters typically
contain more mass in gas than in stars, accretion
can play an important role in determining the final
stellar masses. Numerical simulations of gas accretion
in stellar clusters have found that the stars compete for the available
reservoir of gas. 
The accretion rates are highly nonuniform and are
determined primarily by each star's position in the 
cluster.  Stars in the centre accrete more gas, 
resulting in initial mass segregation.  This 
competitive accretion naturally results in a mass 
spectrum and is potentially the dominant mechanism for
producing the initial mass function.  Furthermore,
accretion on to the core of a cluster forces it to
shrink, which may result in formation of massive stars through
collisions. 

\end{abstract}


\section{Introduction}

One of the most important goals of a general theory of star formation
is to explain the origin of the initial mass function (IMF).  In order
to do this, we need to understand the differences between low-mass and
high-mass star formation. A stellar cluster is the natural size-scale
to investigate these differences as they contain the full mass range of stars.
In this paper, we review how competitive accretion in clusters can
form the basis of a theory for the IMF.
Competitive accretion arises when a group of stars compete for a
finite mass-reservoir (Zinnecker~1982). If this accretion contributes
a large fraction of the final stellar mass, then the competition
process  determines the overall distribution of stellar masses.

Surveys of star forming regions have found that the majority of
pre-main sequence stars are found in clusters (e.g Lada et. al. 1991;
Lada, Strom \& Myers~1993; see also Clarke, Bonnell \&
Hillenbrand~2000).  The fraction of stars in clusters depends on the
molecular cloud considered but generally varies from 50 to $\simgreat
90$ per cent. These clusters  contain
anywhere from tens to thousands of stars with typical numbers of
around a hundred (Lada et. al.~1991; Phelps \& Lada~1997; Clarke
et. al.~2000). Cluster radii are generally a few tenths of a
parsec such that mean stellar densities are of the order of $\approx
10^3$ stars/pc$^3$ (c.f. Clarke et. al.~2000) with central stellar
densities of the larger clusters (e.g. the ONC) being $\simgreat 10^4$
stars/pc$^3$ (McCaughrean \& Stauffer~1994; Hillenbrand \&
Hartmann~1998; Carpenter et. al.~1997).

Furthermore, young clusters are usually associated with massive clumps
of molecular gas (Lada~1992). Generally, the mass of the gas in the
youngest clusters is larger than that in stars (Lada~1991), with up to
90 \% of the cluster mass in the form of gas.  Gas can thus play an
important role in the dynamics of the clusters and affect the final
stellar masses through accretion.

Surveys of the stellar content of young (ages $\approx 10^6$ years)
clusters (e.g. Hillenbrand~1997) reveal that they contain both
low-mass and high-mass stars in proportion as you would expect from a
field-star IMF (Hillenbrand~1997).  Furthermore, there is a degree of
mass segregation present in the clusters with the most massive stars
generally found in the cluster cores.

\section{Mass Segregation}

Young stellar clusters are commonly found to have their most massive stars
in or near the centre (Hillenbrand~1997; Carpenter \etal~1997). 
This mass segregation is similar to that found in older clusters but
the young dynamical age of these systems offers the chance to test
whether the mass segregation is an initial condition or due to the
subsequent evolution.
We know that two-body relaxation drives a stellar system towards
equipartition of kinetic energy and thus towards mass segregation.  In
gravitational interactions, the massive stars tend to lose some of
their kinetic energies to lower-mass stars and thus sink to the centre
of the cluster.

Numerical simulations of  two-body relaxation have shown
that while some degree of mass segregation can occur over the short lifetimes
of these young clusters, it is not sufficient to explain the observations
(Bonnell \& Davies~1998).
Thus the observed positions of the massive stars near the centre of clusters like the ONC reflects the initial conditions of the cluster and of massive star
formation that occurs preferentially in the centre of rich clusters.

Forming massive stars in the centre of clusters is not straightforward
due to the high stellar density. For a star to fragment out of the
general cloud requires that the Jeans radius, the minimum radius for the
fragment to be gravitationally bound, 
\begin{equation} 
R_J \propto  T^{1/2} \rho^{-1/2}, 
\end{equation}
be less than the stellar
separation. This implies that the gas density has to be high,
as you would expect at the centre of the cluster potential. The difficulty 
arises in that the high gas density implies that the fragment mass, being
approximately the Jeans mass, 
\begin{equation}
\label{Jeans_mass}
M_J \propto T^{3/2} \rho^{-1/2}, 
\end{equation} 
is quite low. Thus, unless the
temperature is unreasonably high in the centre of the cluster {\bf
before} fragmentation, the initial stellar mass is quite low.
Equation~(2) implies that the stars in the centre of the cluster
should have the lowest masses, in direct contradiction with the
observations. Therefore, we need a better explanation for the
origin of massive stars in the centre of clusters.

\section{The dynamics of accretion in clusters}

\begin{figure}
\vspace{-0.05truein}
\centerline{\psfig{{figure=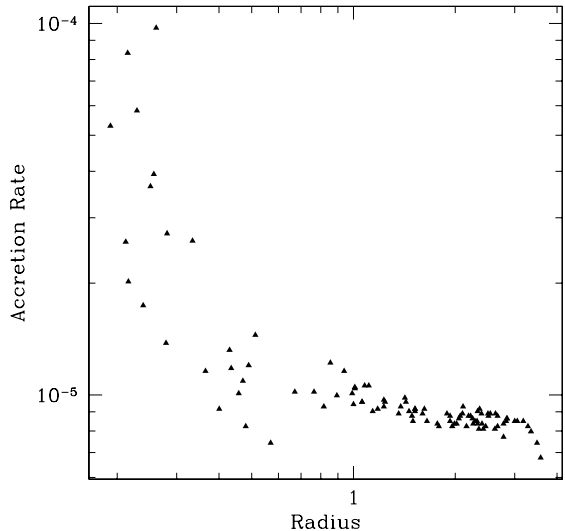,width=3.250truein,height=3.250truein,rwidth=3.25truein,rheight=3.0truein}}}
\caption{\label{cluacc100} The accretion rates versus radius for a cluster
containing 100 stars and 90 \% of its mass in gas.}
\end{figure}

Young stellar clusters are commonly found to be gas-rich with
typically 50 \% to 90 \% of their total mass in the form
of gas (e.g. Lada~1991). This
gas can interact with, and be accreted by, the stars as 
both move in the cluster. If significant accretion occurs, it can affect
both the dynamics and the masses of the individual stars (e.g. Larson~1992). 

Fragmentation models of multiple systems and of stellar clusters show
that the fragmentation is inherently inefficient with a small fraction
of the total mass in the initial fragments (eg. Larson~1978;
Boss~1986; Bonnell \etal~1992; Boss~1996; Klessen, Burkert \&
Bate~1998).  The remaining gas is accreted by the fragments on the gas
free-fall timescale. This occurs as the gas is self-gravitating and is
the dominant mass component.  The free-fall timescale is roughly the
crossing or dynamical time of a stellar cluster as both are related to
the total mass in the cluster which is mostly in the form of gas. Thus
the gas is accreted on the same timescale on which the stars move.  An
exception to this is if the pre-fragmented cluster is highly
structured (e.g. Klessen \etal~1998), then the initial dynamical time
can be significantly longer than the local gas free-fall
timescale. Such clusters should have little gas by the time they have
relaxed to a quasi-spherical distribution. Thus, for the remainder of
this paper, we assume that the initial cluster is approximately
spherical and that the gas and stars have similar distributions.

Simulations of accretion in clusters have been performed for clusters
of 10 to 100 stars using a combination SPH and N-body code. These
simulations found that accretion is a highly non-uniform process where
a few stars accrete significantly more than the rest (Bonnell
\etal~1997; Bonnell \etal~2000).  Individual stars' accretion rates
depend primarily on their position in the cluster (see
Fig.~\ref{cluacc100}) with those in the centre accreting more gas than
those near the outside. Stars near the centre accrete more gas than do
others further out due to the effect of the cluster potential which
funnels the gas down towards the deepest part of the potential. The
accretion rates can also be relatively large when the gas is the
dominant component such that the final masses of the more massive
stars are due to the accretion process. In contrast, many of the stars
do not accrete significant amounts of gas and their final masses are a
closer reflexion of any initial mass distribution in the cluster due
to the fragmentation.

Accretion in stellar clusters naturally leads to both a mass spectrum
and mass segregation. Even from initially equal stellar masses, the
competitive accretion results in a wide range of masses with the most
massive stars located in or near the centre of the
cluster. Furthermore, if the initial gas mass-fraction in clusters is
generally equal, then larger clusters will produce higher-mass stars
and a larger range of stellar masses as the competitive accretion
process will have more gas to feed the few stars that accrete the most
gas.

\begin{figure}
%\vspace{-0.1truein}
\psfig{{figure=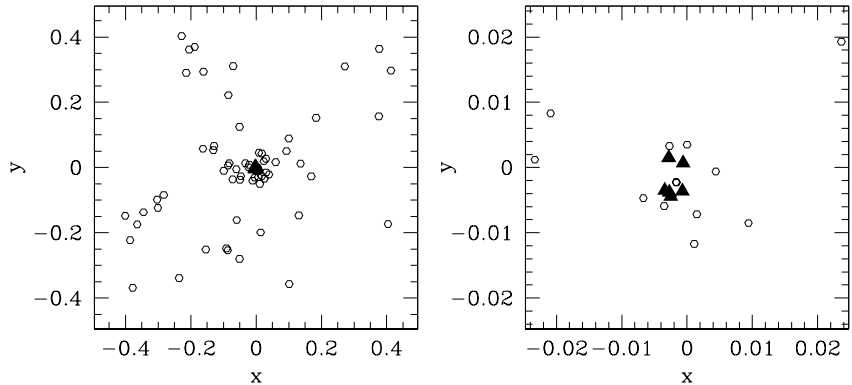,width=4.7truein,height=4.7truein,rwidth=4.7truein,rheight=2.3truein}}
\caption{\label{accmassseg} The position of the 6 most massive stars
(with $M_{\star} \simgreat 4 M_{\rm med}$, filled triangles) in a cluster of
100 stars. The right-hand panel is a blow-up of the cluster core.
The masses of the most massive stars are due to the
competitive accretion process.}
\end{figure}

\section{Modelling Competitive Accretion}

  The accretion  process outlined above is termed ``competitive
accretion'' (Zinnecker~1982) as each star competes for the available
gas reservoir. In order to investigate the possible mass functions
from this process, we need to consider larger clusters, or numbers of clusters,
to get statistically significant numbers. This cannot be done by the
above simulations as numerical resolution is presently inadequate to follow
the accretion process with more than 100 stars. One option is an analytical
or semi-analytical approach to competitive accretion which can then
be applied to larger clusters (Bonnell \etal~2000). 

Gas accretion by a star is given by the general formula
\be 
\macc \approx \pi \rho \vel \racc^2,
\ee
where $\rho$ is the gas density and $\vel$ is the relative gas-star
velocity and $\racc$ is the accretion radius. In a cluster model,
we have the velocities and densities.
Thus, in order to parametrise the accretion process, we need a
description of the accretion radius $\racc$.

Accretion by a star was first explored by Bondi and Hoyle (Bondi \&
Hoyle~1944; Bondi~1952) in terms of an isolated star in a uniform,
non-self-gravitating medium. In this model, the accretion radius is
given by
\begin{equation}
\rbh = { G\ms \over (\vel^2 + c_s^2)^{1/2}},
\ee
where $\ms$ is the stellar mass, and $c_s$ is the gas sound speed.
This approach
neglects the self-gravity of the gas,  the presence of other stars, 
the cluster potential and how these affect the accretion.

An alternative to Bondi-Hoyle accretion is to consider a tidal
accretion radius where gas can only be accreted  onto a star if
it is more bound to it then to other stars or the cluster as a whole.
Accretion in this context is then similar to the Roche-lobe overflow
problem. Taking the Roche-lobe radius to be the accretion radius we have
\be
\rroche = 0.5 \large({\ms\over \menc}\large)^{1/3} \rs,
\ee
where $\menc$ is the mass enclosed in the cluster at the star's
position $\rs$. This approach is consistent with the tidal
effects of the cluster but does not consider whether the gas
is bound to the star when considering it's thermal and kinetic energies.

Support for using such a model comes from
studies of accretion in binary systems (Bate~1997, Bate \& Bonnell~1997).
These studies 
found that the accretion of cold gas was well represented when the accretion
radius was taken to be the Roche-lobe of the individual stars.

The simulations of accretion in clusters were used to test which
model best represented the accretion process (Bonnell \etal~2000).
Figure~\ref{cluacc30} shows a comparison of the SPH determined accretion
rate versus that estimated from Bondi-Hoyle and from Roche-lobe accretion
for a cluster of 30 stars embedded in cold gas. We see that the
Bondi-Hoyle accretion is too high early on in the evolution when
the SPH accretion rate is low and that overall there is little correspondence
between the Bondi-Hoyle accretion rate which is nearly constant and the
SPH determined accretion rate. In contrast, the Roche-lobe accretion 
follows the SPH determined accretion rate from the early low values to
the much higher values that occur towards the end of the simulation.

\begin{figure}
%\vspace{-0.1truein}
\centerline{\psfig{{figure=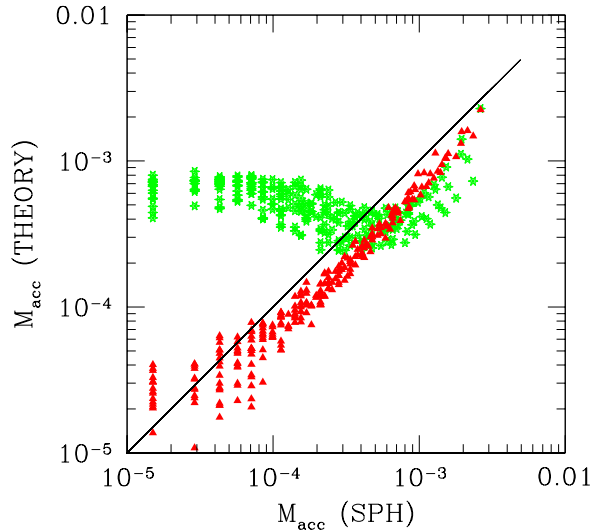,width=3.2truein,height=3.2truein}}}
\caption{\label{cluacc30} Estimates of the accretion rate
for a cluster of 30 stars is plotted for Bondi-Hoyle accretion (stars)
and Roche-lobe accretion (triangles) against the SPH determined accretion 
rates. The gas initially comprises 82 \% of the cluster mass and is cold.}
\end{figure}

That the Roche-lobe accretion works better when the Bondi-Hoyle
accretion gives a higher accretion rate makes sense as the Bondi-Hoyle
radius is then larger than the Roche-lobe radius and the effective
accretion radius would be the minimum of the two. In contrast, it is
surprising (at first) that the Roche-lobe accretion works better even
when the Bondi-Hoyle radius is smaller than the Roche radius. After
closer inspection, it is apparent that the star is  carrying an envelope
of gas with it through the cluster and that this envelope approximately
fills the Roche-lobe. This envelope forms while the stars are initially
moving subsonically and subsequently acts to dampen the relative
high velocity gas so that it can be bound to the star. Simulations where
the stars are initially moving supersonically and are devoid of an envelope
are found to be less well modelled by Roche-lobe accretion, in agreement
with this interpretation. In general, we expect that all stars will
form with circumstellar envelopes and thus the Roche-lobe accretion
should be a good estimate of the accretion rates.

\section{Accretion and the IMF}

Using the above formulation for Roche-lobe accretion in a stellar cluster,
we can estimate plausible IMFs considering simple cluster models.
Hillenbrand \& Hartmann~(1998) showed that the ONC can be
approximated by a King model which has a stellar density profile
$n \propto r^{-2}$. In this case, the number of stars at a given
radius is constant, $dn \propto dr$. The gas density is somewhat
more tricky but two possible distributions are $\rho \propto r^{-2}$,
similar to the stellar distribution, or $\rho \propto r^{-3/2}$, corresponding
to an accretion solution (e.g. Shu~1977; Foster \& Chevalier~1993).
Considering these two possibilities and that the stellar
velocities are in virial equilibrium with the dominant gas distribution, we
can calculate $\ms(r)$ and thus an IMF. 

\begin{figure}
\vspace{-0.5truein}
\centerline{\psfig{{figure=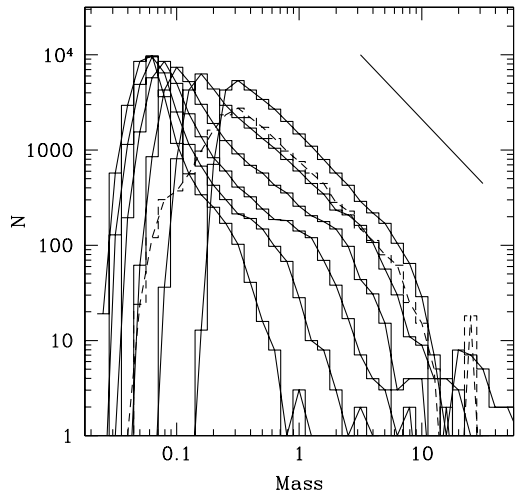,width=3.2truein,height=3.2truein}}}
\caption{\label{cluimf} Mass functions for an analytic model
of competitive accretion in clusters using Roche-lobe accretion. Different
lines show the evolution from an initial distribution containing only
very low-mass stars. The mass functions are summed over 28 clusters
of 1000 stars each. The diagonal line at the right of the graph denotes
 the Salpeter slope of -2.35. The stars do not move in this model.}
\end{figure}

In the first case where $\rho\propto r^{-2}$, we find that $\ms \propto r^{-2}$
and that the resulting mass function is 
\be
dn \propto m^{-3/2} dm.
\ee
For the second case where $\rho\propto r^{-3/2}$, then $\ms \propto r^{-3/4}$,
and the IMF is then 
\be
dn \propto m^{-7/3} dm.
\ee
In both cases, we see that we expect the massive stars to be segregated 
towards the centre of the cluster. 

A plausible evolutionary picture would have the gas density evolving
from a $\rho\propto r^{-2}$ initial distribution to a $\rho\propto
r^{-3/2}$ as material is depleted due to the accretion and is replaced
with gas inflowing from larger radii. As this occurs from the inside
out, stars near the centre of the cluster, which are the more massive
stars, would have the steeper mass-profile. An illustration of the
type of mass function that results from this process is shown in
figure~\ref{cluimf}. Limitations of this approach is that it neglects
the stellar dynamics of the cluster and assumes a simple
prescription for the gas and thus for the stellar velocities.

\begin{figure}
\vspace{-0.5truein}
\centerline{\psfig{{figure=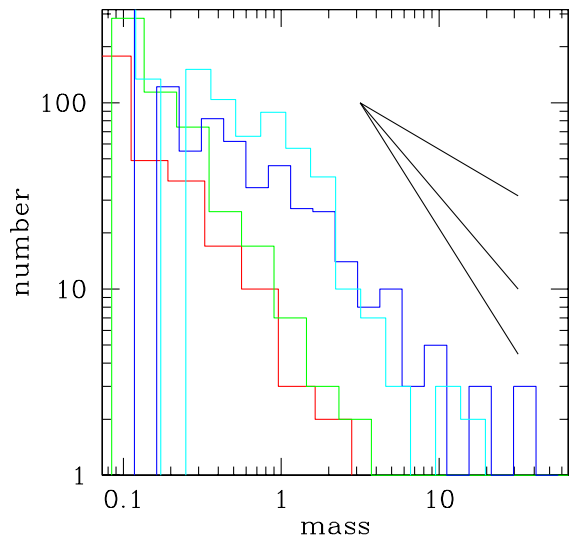,width=3.2truein,height=3.2truein}}}
\caption{\label{cluimf1k} Mass functions for simulated
clusters of 1000 stars accreting gas using a Roche-lobe accretion
approximation. The gas is initially 90 per cent of the cluster mass
and the different curves represent different thermal temperatures
of the gas (cold, virialised) and either an isothermal or adiabatic
equation of state. The three diagonal lines at the right of the graph denote
slopes of -1.5, -2 and the Salpeter slope of -2.35.}
\end{figure}

An alternative is to follow both the stellar and gas-dynamics
simultaneously but with a prescription of the accretion to lessen
resolution requirements (Bonnell \etal~2000).  Figure~\ref{cluimf1k}
shows the results of four clusters of 1000 stars each undergoing
competitive accretion using the Roche-lobe formalism.  All four
clusters result in reasonable IMFs with exponents in the range of
$-1.5$ to $\approx -2.5$ where the Salpeter IMF is $-2.35$. The four
models shown in figure~\ref{cluimf1k} span the range of possible gas
temperatures (virialised, cold) and equations of state (isothermal,
adiabatic). The one major limitation for these models is the lack
of any feedback from the stars and the possibility of turbulence in the gas.

From these models we see that competitive accretion gives reasonable
IMFs and fulfills a basic requirement of producing the more massive stars
near the centre of the cluster. Unfortunately, there is an added complication
in forming very massive stars, $\ms \simgreat 10 \solm$.

\section{Formation of Massive Stars}

The formation of massive stars is problematic not only for their
special location in the cluster centre, but also due to the fact that
the radiation pressure from massive stars is sufficient to halt the
infall and accretion (Yorke \& Krugel~1977; Yorke~1993). This occurs
for stars of mass $\simgreat 10$ \solmas. 

A secondary effect of accretion in clusters is that it can force it to
contract significantly. The added mass increases the binding energy of
the cluster while accretion of basically zero momentum matter will
remove kinetic energy.  If the core is sufficiently small that its
crossing time is relatively short compared to the accretion timescale,
then the core, initially at $n\approx 10^4$ stars pc$^{-3}$, can
contract to the point where, at $n\approx 10^8$ stars pc$^{-3}$,
stellar collisions are significant (Bonnell, Bate \& Zinnecker~1998).
Collisions between intermediate mass stars ($2 \solm \simless m
\simless 10 \solm$), whose mass has been accumulated through accretion
in the cluster core, can then result in the formation of massive ($m
\simgreat 50 \solm$) stars. This model for the formation of massive
stars predicts that the massive stars have to be significantly younger
than the mean stellar age due to the time required for the core to
contract (Bonnell \etal~1998).

Preliminary studies of possible IMFs that would result from a merger
process (Bailey~1999) show that, plotted as a cumulative distribution
to lessen the effects of small statistics, the mass function is compatible
with the high mass function of Kroupa, Tout \& Gilmore~(1990) where
$dn \propto m^{-2.5} dm$.

\section{Summary}

Competitive accretion in young, gas rich stellar clusters is an
appealing mechanism to explain the origin of the IMF. This one
simple physical process can explain both the initial mass segregation
in stellar clusters and potentially the exact mass-distribution.
Stellar dynamical effects such as two-body relaxation are not
able to explain the mass segregation found in clusters such as the ONC
due to their extreme youth.

The accretion in a cluster environment is found to be better represented
by a tidal or Roche-lobe accretion radius than by Bondi-Hoyle accretion.
This occurs as the tidal radius determines when the gas is bound
to the star compared to the cluster as a whole. Furthermore, this
radius represents the maximum extend of any circumstellar envelope which
can act to sweep up the intracluster gas.
 
Gas
accretion in a stellar cluster is highly competitive and uneven. Stars
near the centre of the cluster accrete at significantly higher rates
due to their position where they are aided in attracting the gas by
the overall cluster potential. This competitive accretion naturally
results in both a spectrum of stellar masses, and an initial mass
segregation even if all the stars originate with equal
masses.

Simple analytical models of the cluster and the competitive accretion
yield IMFs which range from $\gamma=-3/2$ when the gas is assumed
to be in the form $\rho\propto r^{-2}$ to $\gamma = -7/3$ when the gas
is in a $\rho\propto r^{-3/2}$ distribution as in an accretion flow.
These two power-laws could be expected to represent the low-mass
stars in the outer part of the cluster and the higher-mass stars
in the inner parts of the cluster, respectively, as the accretion
flow would grow from the inside-out. Simulations of clusters
undergoing the Roche-lobe prescription for accretion produce mass functions
which are compatible with these limits.

Finally, massive stars may form through stellar collisions in the
centre of dense clusters. The necessary density for collisions would
result due to the accretion process of adding mass without significant
momentum which then forces the core to contract to higher
densities. Such a collisional model for the formation of massive stars
evades the problem of accreting onto massive stars.

\end{document}